\documentclass[twocolumn]{jjap2} 
\usepackage{epsfig}
\usepackage{bm}

\title{Correlation Effects in Quantum Dot Wave Function Imaging}

\author{Massimo \textsc{Rontani}$^{1}$\thanks{E-mail address: rontani@unimore.it} and Elisa \textsc{Molinari}$^{1,2}$}

\inst{$^{1}$INFM-CNR National Research Center on nanoStructures
and bioSystems at Surfaces (S3), Modena, Italy \\
$^{2}$Dipartimento di Fisica, Universit\`a degli Studi di Modena e Reggio Emilia, Via Campi 213/A, 41100 Modena, Italy }

\abst{We demonstrate that in semiconductor quantum dots wave functions,
as imaged by local tunneling spectroscopies like STM,
show characteristic signatures of electron-electron 
Coulomb correlation. We predict that such images correspond to 
``quasi-particle'' wave functions which cannot be computed by standard
mean-field techniques in the strongly correlated regime.
From the configuration-interaction solution
of the few-particle problem for prototype dots,
we find that quasi-particle wavefunction images
may display signatures of Wigner crystallization.}

\kword{STM, tunneling spectroscopies, semiconductor quantum dots, electron solid, configuration interaction}

\begin{document}
\maketitle

\section{Introduction} 

Present single-electron tunneling spectroscopies in semiconductor
quantum dots\cite{Jacak,Bimberg,Reimann} (QDs)
provide spectacular images of QD wave functions.
\cite{Grandidier,Millo,Maltezopoulos,Vdovin,Patane,Wibbelhoff}
The measured intensities are generally identified
with the density of carrier states 
at the resonant tunneling (Fermi) energy,
resolved in either real \cite{Grandidier,Millo,Maltezopoulos} or
reciprocal \cite{Vdovin,Patane,Wibbelhoff} space.
However, Coulomb blockade phenomena and strong 
inter-carrier correlation, which are the fingerprints of 
QD physics, complicate the above simple picture. \cite{brief}
Indeed, QDs can be strongly interacting objects with a
completely discrete energy spectrum, which in turn depends on the
number of electrons,\cite{Jacak,Reimann} $N$.
Therefore, orbitals can be ill-defined,
losing their meaning due to interaction. Also, it is
unclear how many electrons one should take into account
to calculate the local density of states,
as a particle tunnels into a QD and the number of electrons
filling the dot flctuates between $N-1$ and $N$ (such
fluctuation is the origin of either the Coulomb current peak or 
the capacitive signal). \cite{Wibbelhoff,Maan,Seigo,Ashoori} 

Here we clarify the physical quantities actually 
probed by scanning tunneling microscopies
\cite{Grandidier,Millo,Maltezopoulos} (STM)
or magneto-tunneling spectroscopies
\cite{Vdovin,Patane,Wibbelhoff,Maan} of QDs, and
how they depend on interactions. 
If only one many-body state is probed at a time,
then the signal is proportional to the probability density
of the {\em quasi-particle} (QP) being injected into the interacting QD.
We demonstrate that the QP density dramatically
depends on the strength of correlation inside the dot, and
it strongly deviates from the common mean-field
(density functional theory, Hartree-Fock) picture in 
physically relevant regimes.

\section{Theory of Quasi-Particle Imaging}\label{s:theory}

The imaging experiments, in their essence, measure
quantities directly proportional to the
probability for transfer of an electron through a barrier, from an
emitter, where electrons fill in a Fermi sea, to a dot, with
completely discrete energy spectrum.
In multi-terminal setups one can neglect the role of electrodes
other than the emitter, to a first approximation.
The measured quantity can be
the current, \cite{Grandidier,Vdovin} the differential
conductance, \cite{Millo,Patane,Maltezopoulos,Seigo} or the QD
capacitance, \cite{Wibbelhoff,Maan,Ashoori} while the emitter can be the STM
tip, \cite{Grandidier,Millo,Maltezopoulos} or a $n$-doped GaAs
contact, \cite{Vdovin,Patane,Wibbelhoff,Maan,Seigo,Ashoori}
and the barrier can be the
vacuum \cite{Grandidier,Millo,Maltezopoulos} as well as a AlGaAs
spacer. \cite{Vdovin,Patane,Wibbelhoff,Maan,Seigo,Ashoori}

According to the seminal paper by Bardeen, \cite{Bardeen}
the transition probability (at zero temperature) is given by the expression
$(2\pi/\hbar)\left|{\cal{M}}\right|^2n(\epsilon_f)$, where
$\cal{M}$ is the matrix element and $n(\epsilon_f)$ is the energy
density of the final QD states. The common wisdom would predict the
probability to be proportional to the total density of QD states
at the resonant tunneling energy, $\epsilon_f$, possibly space-resolved
since $\cal{M}$ would depend on the resonant QD orbital. \cite{Tersoff}
Let us now assume that: (i) Electrons in the emitter do not interact
and their energy levels form a continuum.
(ii) Electrons from the emitter access through the barrier a single QD at 
a sharp resonant energy, corresponding to a well defined interacting QD state.
(iii) The $xy$ and $z$ motions of electrons are separable, the
$z$ axis being parallel to the tunneling direction. (iv) Electrons
in the QD all occupy the same confined orbital along $z$, 
$\chi_{\text{QD}}(z)$.
Then one can show \cite{brief} that the matrix element $\cal{M}$ may
be factorized as
\begin{equation}
{\cal{M}}\propto T M,
\end{equation}
where $T$ is a purely single-particle matrix element while
the integral $M$ contains the whole correlation physics.

The former term is proportional to the current density evaluated
at any point $z_{\text{bar}}$ in the barrier:
\begin{equation}
T = \frac{\hbar^2}{2m^*}\left[\chi^*_{\text{E}}(z)
\frac{ \partial \chi_{\text{QD}}(z) }{\partial z}
-\chi_{\text{QD}}(z)\frac{\partial \chi_{\text{E}}^*(z) }{\partial z}
\right]_{ z=z_{\text{bar}} },
\label{eq:T}
\end{equation}
where $\chi_{\text{E}}(z)$ is the resonating emitter state along $z$ 
evanescent in the barrier and $m^*$ is the electron effective mass.
The term (\ref{eq:T}) contains the information regarding the
overlap between emitter and QD orbital tails in the barrier,
$\chi_{\text{E}}(z)$ and $\chi_{\text{QD}}(z)$, respectively.
Since $T$ is substantially independent from both $N$ and $xy$ location,
its value is irrelevant in the present context. 

On the other hand, the in-plane matrix element $M$ conveys 
the information related to correlation effects:
\begin{equation}
M = \int \phi^*_{\text{E}}(\bm{\varrho})\,
\varphi_{\text{QD}}(\bm{\varrho})\,\text{d}\,\bm{\varrho},
\label{eq:M}
\end{equation}
where $\varphi_{\text{QD}}(\bm{\varrho})$ is the QP
wavefunction of the interacting QD system: \cite{nota}
\begin{equation}
\varphi_{\text{QD}}(\bm{\varrho}) = \langle N - 1 | \hat{\Psi}(\bm{\varrho})|
N \rangle .
\label{eq:def}
\end{equation}
Here $\phi_{\text{E}}$ is the
in-plane part of the emitter resonant orbital, $\hat{\Psi}(\bm{\varrho})$
is the fermionic field operator destroying an electron
at position $\bm{\varrho}\equiv (x,y)$, $| N - 1 \rangle $
and $| N \rangle $ are the QD interacting ground states
with $N-1$ and $N$ electrons, respectively (see
also Sec.~\ref{s:2DQD}). We omit spin indices for the sake of
simplicity.

Results (\ref{eq:M}-\ref{eq:def}) are the key
for predicting  wave function images both in real
and reciprocal space.
In STM, $\phi_{\text{E}}(\bm{\varrho})$ is the localized tip
wave function; if we ideally assume it point-like and
located at $\bm{\varrho}_0$ \cite{Tersoff},
i.e.~\protect{$\phi_{\text{E}}(\bm{\varrho}) \approx \delta(\bm{\varrho}
-\bm{\varrho}_0)$}, then the signal intensity is proportional
to $\left|\varphi_{\text{QD}}(\bm{\varrho}_0)\right|^2$,
which is the usual result of the one-electron
theory \cite{Maltezopoulos,Tersoff}, provided the ill-defined QD orbital
is replaced by the QP wavefunction unambiguously defined by Eq.~(\ref{eq:def}).
In magneto-tunneling spectroscopy, the emitter in-plane wavefunction
is a plane wave, \protect{$\phi_{\text{E}}(\bm{\varrho})=
\text{e}^{\text{i}\bm{k\cdot\varrho}}$}, and the matrix element
(\ref{eq:M}) is the Fourier transform of $\varphi_{\text{QD}}$,
\protect{$M = \varphi_{\text{QD}}(\bm{k})$}.
Again, we generalize the standard one-electron result \cite{Patane} by
substituting $\varphi_{\text{QD}}(\bm{k})$ for
the QD orbital.
Note that $M$ is the relevant quantity
also for intensities in space-integrated spectroscopies probing the QD
addition energy spectrum \cite{Seigo,Ashoori}.

\section{Two-dimensional Quantum Dot}\label{s:2DQD}

\subsection{The Non-Interacting Case}

We now apply the theory of Sec.~\ref{s:theory} to a two-dimensional
parabolic QD with a few strongly interacting electrons. The harmonic
potential was proven to be an excellent description of several
experimental traps. \cite{Reimann}
The non-interacting effective-mass Hamiltonian of the $i$-th electron is
\begin{equation}
H_{0}(i)\,=\,\frac{p_i^2}{2m^{*}}
+\frac{1}{2}m^*\omega_0^2\varrho_i^2.
\label{eq:HSP}
\end{equation}
The eigenstates $\varphi_{a}(\bm{\varrho})$ of (\ref{eq:HSP})
are known as Fock-Darwin orbitals \cite{jcc}. Their peculiar 
shell structure, with constant energy spacing $\hbar\omega_0$, 
is represented in Fig.~\ref{f:fluct}
up to the third shell.
\begin{halffigure}[h]
\setlength{\unitlength}{1 cm}
\begin{picture}(8.6,4.0)
\put(0.5,0.5){\epsfig{file=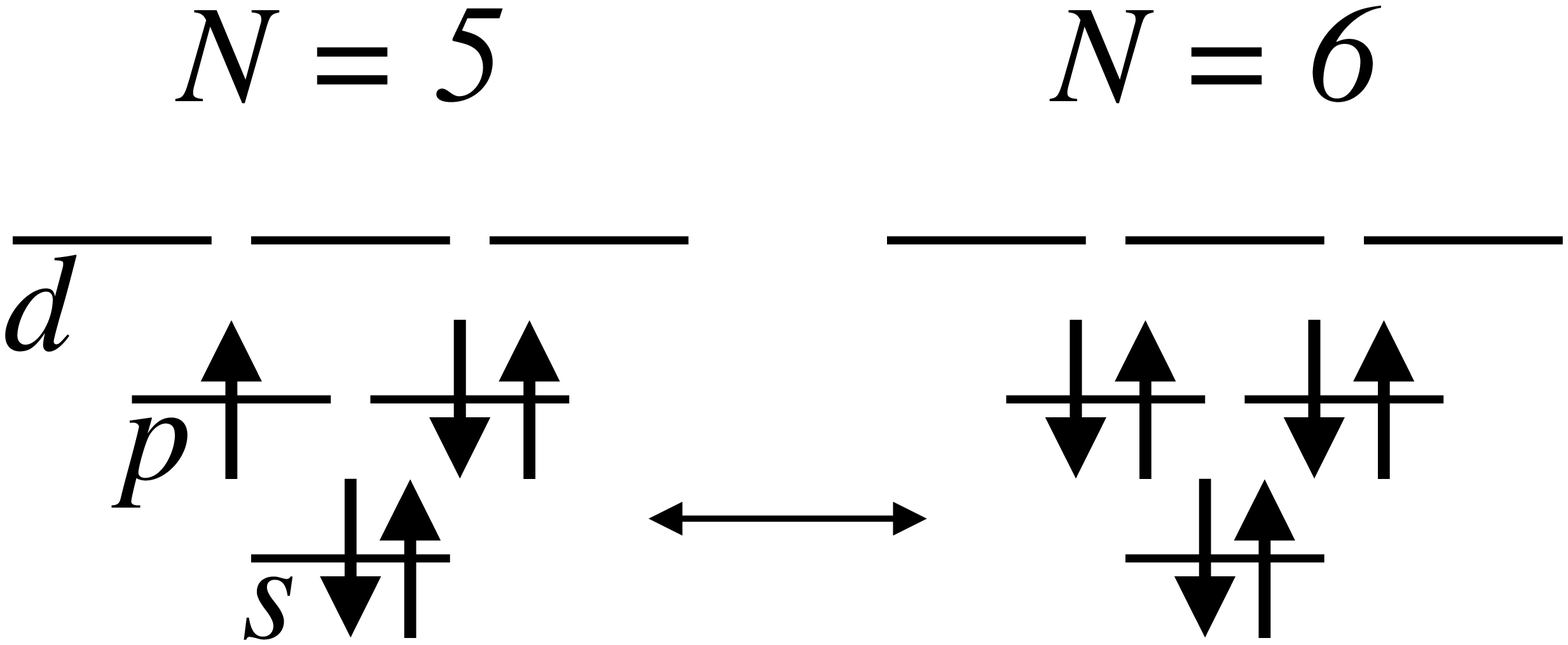,angle=0,width=7.5cm}}
\end{picture}
\caption{Electronic configuration of  the non-interacting
quantum dot ground state as the electron number, $N$, fluctuates
between 5 and 6. The arrows represent electrons (with their spin) filling
Fock-Darwin orbitals. The letters identify different energy shells
(the third-shell central orbital has $s$ character).}
\label{f:fluct}
\end{halffigure}

What is $\varphi_{\text{QD}}(\bm{\varrho})$ in the non-interacting case?
Let us consider e.g.~$N=6$. Then, the 5-electron ground state 
$| N - 1 \rangle $ appearing in the definition (\ref{eq:def}) is naturally 
obtained from the Aufbau principle of atomic physics: \cite{Seigo} 
all and only the lowest-energy Fock-Darwin orbitals are filled in with 
electrons according to Pauli exclusion principle 
(left panel of Fig.~\ref{f:fluct}). The 6-electron ground state 
$| 6 \rangle = \hat{c}_{p\uparrow}^{\dagger} | 5 \rangle $
is obviously obtained from $| 5 \rangle $ by adding a spin-up
electron into the lowest-energy $p$-type empty orbital
(right panel of Fig.~\ref{f:fluct}); here $\hat{c}_{p\uparrow}^{\dagger}$
is the pertinent creation operator.
By expanding the field operator $\hat{\Psi}$ 
on the Fock-Darwin orbital basis,
\protect{$\hat{\Psi}(\bm{\varrho},s_z)=\sum_{a\sigma}\varphi_a(\bm{\varrho})
\,\xi_{\sigma}(s_z)\,{\hat{c}}_{a\sigma}$}
[\protect{$\xi_{\sigma}(s_z)$ is the spin part of the
electron wave function with eigenvalue $\sigma = \uparrow,\downarrow$
and spin coordinate $s_z$}],  we derive that the orbital
part of the QP wave function is simply $\varphi_{\text{QD}}(\bm{\varrho})
= \varphi_p(\bm{\varrho})$. 

The above result is a sensible one: as we inject e.g.~via the STM tip
an additional electron to the 5-electron ground state and $N$
oscillates between 5 and 6 (Fig.~\ref{f:fluct}), the non-interacting
wave function of the tunneling electron can be regarded alternatively 
either as the lowest-energy unoccupied orbital when $N=5$
(Fig.~\ref{f:fluct} left panel) or as the highest-energy occupied
orbital when $N=6$ (Fig.~\ref{f:fluct} right panel).
In the section below we consider the effects of electron-electron interaction.

\subsection{Configuration-Interaction Approach to the Interacting
Problem}

The fully interacting Hamiltonian is the sum of single-particle
terms (\ref{eq:HSP}) plus the Coulomb term:
\begin{equation}
H = \sum_{i=1}^{N}H_{0}(i)+\frac{1}{2}\sum_{i\neq j}\frac{e^{2}}
{\kappa|\bm{\varrho}_{i}-\bm{\varrho}_{j}|},
\label{eq:HI}
\end{equation}
where $\kappa$ is the static relative dielectric constant of the host
semiconductor. We solve numerically the
few-body problem of Eq.~(\ref{eq:HI}), for the ground state at
different $N$'s, by means of
the configuration interaction (CI) method \cite{guy,cpc,jcc}, 
where $|N\rangle$ is expanded in a series of Slater determinants built 
by filling in
a truncated set of Fock-Darwin orbitals with $N$ electrons, and
consistently with symmetry constraints.
From the solution of the resulting large-size matrix-diagonalization
problem, we obtain eigenvalues and eigenvectors of ground-
and first excited-states. \cite{method} 
Then, we evaluate the matrix
element (\ref{eq:def}), by decomposing $|N\rangle$ and
$|N - 1 \rangle$ on the Slater determinant basis: the resulting 
$\varphi_{\text{QD}}(\bm{\varrho})$ is now a mixture of different
Fock-Darwin orbitals, with weights controlled by the strength of
correlation.

\subsection{Tuning the Strength of Correlation}

A way of artificially tuning the strength of Coulomb
correlation in QDs is to dilute the electron density.
While the kinetic energy term scales as $r_s^{-2}$,
$r_s$ being the parameter measuring 
the average distance between electrons, 
the Coulomb energy scales as $r_s^{-1}$.
Therefore, at low enough density,
electrons pass from a ``liquid'' phase,
where low-energy motion is equally controlled by kinetic and
Coulomb energy, to a ``crystallized'' phase, reminescent of the
Wigner crystal in the bulk, where electrons are localized in space
and arrange themselves in a geometrically ordered configuration
such that electrostatic repulsion is minimized. \cite{Reimann}
In the latter regime Coulomb correlation severely mixes many
different Slater determinants, and the CI approach is the ideal 
tool to quantitatively predict correlation effects. \cite{jcc}

The typical QD lateral extension is given by the characteristic
dot radius $\ell_{\text{QD}} = (\hbar/m^*\omega_0)^{1/2}$,
$\ell_{\text{QD}}$ being the
mean square root of $\varrho$ on the Fock-Darwin lowest-energy level 
$\varphi_{s}$.
As we keep $N$ fixed and increase $\ell_{\text{QD}}$, the Coulomb-to-kinetic
energy ratio $\lambda = \ell_{\text{QD}}/a^*_{\text{B}}$ [$a^*_{\text{B}}
=\hbar^2\kappa/(m^*e^2)$
is the effective Bohr radius of the dot] \cite{Egger} increases
as well, driving the system into the ``Wigner'' regime \cite{Brueckner}.
As a rough indication, consider that for $\lambda \approx 2$ or lower
the electronic ground state is liquid, while above $\lambda \approx 4$
electrons form a ``crystallized'' phase. \cite{Egger}

\section{From high to low electron density: Wigner crystallization}

\begin{halffigure}[h]
\setlength{\unitlength}{1 cm}
\begin{picture}(8.6,16)
\put(0.5,0){\epsfig{file=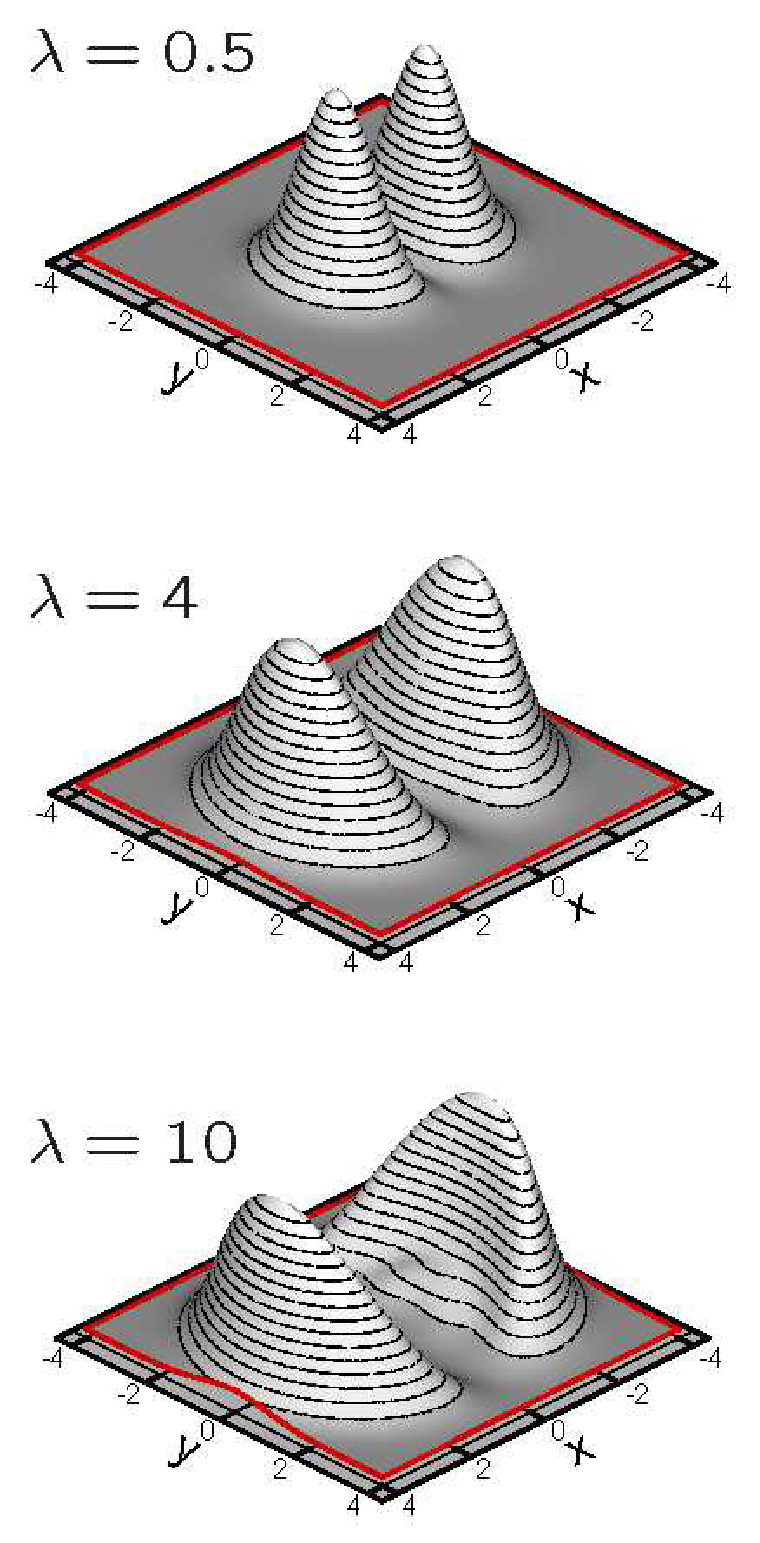,angle=0,width=7.5cm}}
\end{picture}
\caption{Square modulus of the quasi-particle wave function in the quantum
dot plane for three different values of the dimensionless parameter
$\lambda$. This quantity is proportional to the STM signal when
the electron number $N$ fluctuates between 5 and 6. As $\lambda$ increases
(from top to bottom) the density decreases and the wave function shape
evolves from a characteristic $p$-type Fock-Darwin orbital (top panel,
$\lambda=0.5$) into a complex figure peculiar of the ``crystallized'' phase
(bottom panel, $\lambda=10$). The wave function normalization is
arbitrary and the length unit is the characteristic dot radius
$\ell_{\text{QD}}$.
}
\label{f:wow}
\end{halffigure}

Figure \ref{f:wow} shows the square modulus of the QP wave function,
corresponding to the injection of the 6-th electron,
in the $xy$ plane for three different values of $\lambda$.
As $\lambda$ increases (from top to bottom), the density decreases
going from the non-interacting
limit (Fig.~\ref{f:wow}, top panel, $\lambda=0.5$),
deep into the Wigner regime (Fig.~\ref{f:wow}, bottom panel,
$\lambda=10$). At high density ($\lambda=0.5$, approximately corresponding
to the electron density $n_e=3.8\times 10^{12}$ cm$^{-2}$) 
the wave function substantially
coincides with the non-interacting Fock-Darwin $p$ orbital
$\varphi_p(\bm{\varrho})$ of Fig.~\ref{f:fluct}.
By increasing the QD radius (and $\lambda$), the QP wave function
weight moves towards larger values of $\varrho$. 
By measuring lengths in units of $\ell_{\text{QD}}$, as it is done in 
Fig.~\ref{f:wow}, this trivial effect should be totally 
compensated. However, we see in the middle panel of
Fig.~\ref{f:wow} ($\lambda=4$, $n_e \approx 1.5\times 10^{10}$
cm$^{-2}$) that the now much stronger
correlation is responsible for an unexpected weight reorganization,
which is related to the formation of an outer ``ring'' of crystallized
electrons in the Wigner molecule. \cite{Egger}
Such tendency is clearly confirmed at even lower densities 
($\lambda=10$, $n_e \approx 1.3\times 10^9$ cm$^{-2}$). Now, together
with the outer ring, a new structure is visible close to the QD center
(if $\varrho\rightarrow 0$ then $\varphi_{\text{QP}}\rightarrow 0$
due to the orbital $p$ symmetry). Such complex shape is consistent
with the onset of a solid phase with 5 electrons sitting at the
apices of a regular pentagon plus one electron at the center.
\cite{Egger,Bolton,EPLus}

In Fig.~\ref{f:wow} the absolute QP weight has been arbitrarily
renormalized, which we believe to be a sensible procedure in view
of comparison with experimental images.
\begin{halffigure}[h]
\setlength{\unitlength}{1 cm}
\begin{picture}(8.6,6.5)
\put(0.0,0.3){\epsfig{file=f2.eps,angle=0,width=8.3cm}}
\end{picture}
\caption{Quasi-particle wave function vs. $x$ ($y=0$) for different
$\lambda$-values. We use the same parameters as in Fig.~\ref{f:wow}
except that now the wave function normalization is absolute.
As $\lambda$ increases the total weight decreases from 0.97
($\lambda=0.5$) up to 0.15 ($\lambda=10$).
}
\label{f:cut}
\end{halffigure}
In order to illustrate a second correlation effect, in addition
to shape changes, in Fig.~\ref{f:cut} we plot the absolute value
of the QP wave function as a function of $x$ at $y=0$, for the same values
of $\lambda$ as in Fig.~\ref{f:wow}. \cite{brief} Figure \ref{f:cut} clearly
demonstrates a dramatic weight loss as $\lambda$ is increased:
the stronger the correlation, the more effective the orthogonality
between interacting states. Note also in Fig.~\ref{f:cut}
that the shoulder of the outer QP peak close
to the QD center is clearly visible for $\lambda=10$.

\section{Present Status of Imaging Experiments}

Among the existing imaging tunneling experiments, 
\cite{Grandidier,Millo,Maltezopoulos,Patane,Wibbelhoff,Maan}
two have specifically focused on results in the presence
of several carriers in the QD.
\cite{Wibbelhoff,Maan}

A first experiment \cite{Wibbelhoff}
concerned electrons in InAs self-assembled QDs
in the non interacting high-density limit ($\lambda\approx 0.5$).
This work demonstrates the experimental imaging
of an Aufbau-like filling sequence as up to six electrons are 
sequentially injected into the QDs. In particular, the specific 
filling sequence can be understood in terms of the consecutive
filling of the $s$ orbital first, then one of the
two $p$ orbitals of the second shell
(Fig.~\ref{f:fluct}), and eventually the other one.
The above sequence differs from that expected according to Hund's rule,
namely the third and fourth electrons should separately fill in the two 
$p$ orbitals with parallel spins \cite{Seigo}. 
The deviation from the above rule 
is attributed to either piezoelectric effects
or to a slight elongation of the InAs island shape. \cite{Wibbelhoff}
However, many QDs were probed at once and the above interpretation
cannot be regarded as definitive.

In a second experiment the imaging of InAs QD hole wave functions
was addressed. \cite{Maan} The authors observe an anomalous filling
sequence up to 6 holes ($s$, $s$, $p$, $p$, $d$, $d$) and interpret it
in terms of a generalized Hund's rule for the two $p$ and the two $d$ 
orbitals together, namely the total spin should be maximized as $N$ 
increases as an effect of strong Coulomb correlation.
Assuming reasonable hole parameters as $\kappa=12.4$,
$\hbar\omega_0=25$ meV, $m^*=0.3m_e$, we estimate, within
the simple parabolic potential model, the key parameter $\lambda$
to be 1.46.
Such value is comparable to those of typical devices \cite{Seigo}
showing ``standard'' Aufbau physics and it seems to us too small to
support claims of qualitatively new correlation effects. \cite{jcc}
An alternative explanation of the filling sequence could be 
related to merely single particle effects arising from the
complex hole band structure. Specifically, assuming orbital energies 
to be ordered as $s$, $p$, $d$ and no shell degeneracy, then 
electrons should consecutively fill
in such orbitals and Hund's rule would never hold.
Such interpretation seems to be confirmed by
independent theoretical work. \cite{zunger}
Further measurements, as a function
of the magnetic field parallel to $z$,
could be useful to clarify the question.

From the above discussion it appears that experimental
investigations of QP wave functions in regimes where 
correlation effects are significant are lacking so far.
We hope that our results will
stimulate further experiments.

\section{Conclusions}

In conclusion, we have shown that QP wave functions of QDs
are extremely sensitive to electron-electron correlation, and may differ
from single-particle states in physically relevant cases. This result is
of interest to predict the real- and reciprocal-space wave function images
obtained by tunneling spectroscopies, as well as the intensities of
addition spectra of QDs. We believe that our findings will be
important also for other strongly confined systems,
like e.g.~nanostructures at surfaces \cite{Leo}.

\section*{Acknowledgment}

We thank A. Lorke and S. Heun for valuable discussions.
This paper is supported by MIUR-FIRB RBAU01ZEML, MIUR-COFIN 2003020984,
I. T. INFM Calc. Par. 2005, 
Italian Minister of Foreign Affairs,
General Bureau for Cultural Promotion and Cooperation.


\begin{thebibliography}{99} 
\bibitem{Jacak}
L. Jacak, P. Hawrylak and A. W\'ojs: {\em Quantum dots},
(Springer, Berlin, 1998).
\bibitem{Bimberg}
D. Bimberg, M. Grundmann and N. N. Ledentsov:
{\em Quantum dot heterostructures,} (Wiley, New York, 1999).
\bibitem{Reimann}
S. M. Reimann and M. Manninen: Rev. Mod. Phys. {\bf 74} (2002) 1283.
\bibitem{Grandidier}
B. Grandidier, Y. M. Niquet, B. Legrand, J. P. Nys, C. Priester, 
D. Sti\'evenard, J. M. G\'erard and V. Thierry-Mieg: 
Phys. Rev. Lett. {\bf 85} (2000) 1068.
\bibitem{Millo}
O. Millo, D. Katz, Y. Cao and U. Banin: 
Phys. Rev. Lett. {\bf 86} (2001) 5751.
\bibitem{Maltezopoulos}
T. Maltezopoulos, A. Bolz, C. Meyer, C. Heyn, W. Hansen, 
M. Morgenstern and R. Wiesendanger: 
Phys. Rev. Lett. {\bf 91} (2003) 196804.
\bibitem{Vdovin}
E. E. Vdovin, A. Levin, A. Patan\`e, L. Eaves, P. C. Main, 
Yu. N. Khanin, Yu. V. Dubrovskii, M. Henini, and G. Hill: 
Science {\bf 290} (2000) 122.
\bibitem{Patane}
A. Patan\`e, R. J. A. Hill, L. Eaves, P. C. Main, M. Henini,
M. L. Zambrano, A. Levin, N. Mori, C. Hamaguchi, Yu. V. Dubrovskii,
E. E. Vdovin, D. G. Austing, S. Tarucha and G. Hill: 
Phys. Rev. B {\bf 65} (2002) 165308.
\bibitem{Wibbelhoff}
O. S. Wibbelhoff, A. Lorke, D. Reuter and A. D. Wieck: 
Appl. Phys. Lett. {\bf 86} (2005) 092104.
\bibitem{brief}
M. Rontani and E. Molinari: Phys. Rev. B {\bf 71} (2005) 233106.
\bibitem{Maan}
D. Reuter, P. Kailuweit, A. D. Wieck, U. Zeitler, O. Wibbelhoff,
C. Meier, A. Lorke and J. C. Maan: Phys. Rev. Lett. {\bf 94} (2005) 026808,
and private communication.
\bibitem{Seigo}
S. Tarucha, D. G. Austing, T. Honda, R. J. van der Hage and L. P. Kouwenhoven: 
Phys. Rev. Lett. {\bf 77} (1996) 3613.
\bibitem{Ashoori}
R. C. Ashoori: Nature {\bf 379} (1996) 413.
\bibitem{Bardeen}
J. Bardeen: Phys. Rev. Lett. {\bf 6} (1961) 57.
\bibitem{Tersoff}
J. Tersoff and D. R. Hamann: Phys. Rev. B {\bf 31} (1985) 805;
J. Tersoff: Phys. Rev. Lett. {\bf 57} (1986) 440.
\bibitem{nota}
The quantity is also known as the spectral density amplitude of
the one-electron propagator resolved in real space. For analogous treatments
in many-body tunneling theory see
e.g. J. A. Appelbaum and W. F. Brinkman: Phys. Rev. {\bf 186} (1969)
464; T. E. Feuchtwang: Phys. Rev. B {\bf 10} (1974) 4121, and
refs.~therein.
\bibitem{jcc}
M. Rontani, C. Cavazzoni, D. Bellucci and G. Goldoni:
available at cond/mat (2005).
\bibitem{guy}
M. Rontani, S. Amaha, K. Muraki, F. Manghi, E. Molinari, S. Tarucha,
and D. G. Austing: Phys. Rev. B {\bf 69} (2004) 85327.
\bibitem{cpc}
M. Rontani, C. Cavazzoni and G. Goldoni: Comp. Phys. Commun. {\bf 169}
(2005) 430.
\bibitem{method}
Here we implemented a parallel
version of our CI code, allowing for using a Fock-Darwin basis set
as large as 36 orbitals,
and for diagonalizing matrices of linear dimensions
up to $\approx 10^6$. As a convergence test, \cite{jcc} we could accurately
reproduce QMC ground state energies up to $\lambda=10$ and
$N=6$.\cite{Egger}
\bibitem{Egger}
R. Egger, W. H\"ausler, C. H. Mak and H. Grabert:
Phys. Rev. Lett. {\bf 82} (1999) 3320.
\bibitem{Brueckner}
The dimensionless ratio $\lambda$ is the QD analog to the density parameter
$r_s$ in extended systems.
\bibitem{Bolton}
F. Bolton and U. R\"ossler: Superlatt. Microstruct. {\bf 13} (1993) 
139; V. M. Bedanov and F. M. Peeters: Phys. Rev. B {\bf 49} (1994) 2667.
\bibitem{EPLus}
M. Rontani, G. Goldoni, F. Manghi and E. Molinari: 
Europhys. Lett. {\bf 58} (2002) 555.
\bibitem{zunger}
L. He, G. Bester and A. Zunger: cond-mat/0505330.
\bibitem{Leo}
See e.g. P. Jarillo-Herrero, S. Sapmaz, C. Dekker, L. P. Kouwenhoven
and H. S. J. van der Zant: Nature {\bf 429} (2004) 389.
\end{thebibliography}
\end{document}